\documentclass{article}

\newcommand{\iec}{\mbox{i.\,e.\,}}

\newcommand{\egc}{\mbox{e.\,g.\,}}

\newcommand{\dr}[1]{\ensuremath{\mathrm{d} #1\,}}

\newcommand{\ket}[1]{\ensuremath{\left|  #1 \right\rangle}}
\newcommand{\nrm}{\frac{1} {\sqrt{2} } }

\newcommand{\be}{\begin{equation}}
\newcommand{\ee}{\end{equation}}
\newcommand{\e}[1]{\mathrm{e}^{#1}}

\begin{document}
\title{Life and death in the tails of the GRW wave function}
\author{David Wallace\thanks{Department of History and Philosophy of Science/Department of Philosophy, University of Pittsburgh; \texttt{david.wallace@pitt.edu}}}

\date{\today}

\maketitle

\begin{abstract}
It is often assumed that the only effect of the Ghirardi-Rimini-Weber (`GRW')
dynamical collapse mechanism on the `tails' of the wavefunction (that
is, the components of superpositions on which the collapse is \emph{not}
centred) is to reduce their weight. In consequence the tails are often thought to behave
exactly as do the various branches in the Everett interpretation except
for their much lower weight. 

These assumptions are demonstrably
inaccurate: the collapse mechanism has substantial and detectable
effects within the tails. The relevance of this misconception for the
dynamical-collapse theories is debatable, though.
\end{abstract}

(This is a lightly-revised version of a paper first published online in 2014. Related observations about dynamical collapse theories are made by \cite{vaidmandeterminism}; the arguments in the first version of this paper are discussed by \cite{sebenskillercollapse,mcqueenfourtails}.)

\section{Introduction: the problem of tails}

The GRW dynamical-collapse theory, and its more sophisticated
descendants,\footnote{The GRW theory was originally proposed in \cite{grw}, and was significantly revised by \cite{pearle}. For a comprehensive review, see \cite{bassighirardireview}. For philosophical discussion, see section 6 of \cite{wallaceashgate} and references therein.} set out to solve the measurement problem in perhaps the
most direct way possible: by modifying the normal unitary dynamics so as
to replace the ill-defined `projection-postulate' with a
genuine dynamical process which with very high probability collapses
macroscopic superpositions onto macroscopically definite states.

The `problem of tails' \cite{shimonydesideratum,albertloewer1990} arises because (it is
claimed) the GRW collapse mechanism fails to produce states which
actually are macroscopically definite.

In more detail: recall that the fundamental assumption of the GRW model\footnote{The collapse mechanism proposed by GRW has since 
been superseded in technical
work by more sophisticated variants (primarily so as to address the
problem of identical particles) but to the best of my knowledge nothing
conceptually fundamental depends on this change, so for convenience I
will work with the basic GRW model.} is that
each particle (say, the $i$th particle) has a very small random chance per unit time of collapsing via the process

\be
\psi(x_1,\ldots,x_N) \longrightarrow \e{-(x_i-x_0)^2/2a^2}\psi(x_1,\ldots,x_N)
\ee 
(I omit normalisation). The `collapse centre' $x_0$ is determined
randomly, with its probability of being in a region around some $x$
being equal to the `standard' probability of a position measurement
finding the particle at $x$.

Since a Gaussian vanishes nowhere, obviously the collapse mechanism
cannot localise the wavefunction in any region of configuration space
that it was not already localised in. If a `macroscopically definite'
state is supposed to be localised in the region of configuration space
corresponding to our classical notion of the location of that state,
then we have a problem.\footnote{The problem can be avoided via the "primitive ontology" proposal in \cite{goldsteincommonstructure}, which supplements the wavefunction in GRW with additional properties intended to represent the spatially localised entities; I will not be concerned with this strategy here. (In their terms, I am concerned only with $GRW_0$.)}

The problem actually comes in two flavours (here I follow \cite{wallaceashgate}). The first might be called
the problem of `bare tails', and can be stated for a system consisting
of a single particle: a particle in a Gaussian state is not strictly
located in any finite spatial region at all, and so (it is argued)
cannot be regarded as describing a localised classical particle, no
matter how narrow the Gaussian is. The problem of bare tails has
received extensive discussion in the literature recently (see, for
instance \cite{lewis94,lewis2003,cliftonmonton1,cliftonmonton2,ghirardi98,bassi99,montonmassdensity}).

I will not be concerned much with the bare-tails problem in this paper. I
will make one comment, though: the bare-tails problem is not really
anything to do with the GRW theory, but is a natural consequence  of
unitary Schr\"{o}dinger dynamics. Wave-packets with compact support
cannot be created (they require infinite potential wells); if they were to be created, they
would spread out instantaneously.\footnote{\emph{Instantaneously}, not just at
lightspeed: this isn't an artefact of non-relativistic physics. See
\cite{wallaceconceptualqft} and references therein for more on this topic.} As such, if bare tails are a
problem then they are a problem for any version of quantum
mechanics that takes the wave-function as representing macroscopic ontology (such as
the Everett interpretation).\footnote{I don't think they are a problem,
in fact: rather, they demonstrate that the eigenvector-eigenvalue link
is a hopeless way to understand the ontology of realistic quantum
systems \ldots but this is not the topic of the present paper. (See my discussion in \cite{wallacemodern,wallaceorthodoxy}.)}

The problem of `structured tails', by contrast, is explicitly a problem
restricted to dynamical collapse theories. Recall that in a
Schr\"{o}dinger-cat situation, with a state like
\be \label{cat}\nrm \left( \ket{\mbox{alive cat}}+\ket{\mbox{dead cat}} \right)\ee
ideally what we want dynamical collapse to do is to deliver (half the
time, anyway) the state
\be \ket{\mbox{alive cat}}.\ee But the collapse mechanism doesn't
actually do this. Within $10^{-11}$ seconds or so one of the 
particles in
the cat will undergo collapse, with a 50\% chance of concentrating almost 
all its amplitude onto being in the living cat. Since it is entangled
with the remaining particles  in the cat, this will yield a state
something like
\be \label{collapsecat}\alpha \ket{\mbox{alive cat}}+\beta \ket{\mbox{dead cat}} \ee
where $|\alpha|^2 \gg |\beta|^2$ but $\beta \neq 0$.

Arguably, this doesn't do us much good. The dead-cat part of the state
may have very low weight but it's still just as much part of the state.
And in the GRW theory, there is no \emph{conceptual} connection between
mod-squared amplitude and `probability' or `actuality' or anything: the
connection is supposed to be purely dynamical, manifesting via the
collapse process. So it seems that we still have macroscopic
superpositions, and that dynamical collapse has not after all solved the
measurement problem.

The problem can be sharpened by comparing dynamical-collapse theories to
the Everett interpretation (here I follow \cite{corderogrw}). Modern
versions of the Everett interpretation do not introduce `worlds' or
`minds' as extra terms in the formalism: rather, they make use of
dynamical decoherence to show that the unitarily-evolving wave-function
is a superposition of essentially-independent quasi-classical worlds. In
my preferred form of the interpretation (see \cite{wallacebook} or \cite{wallaceFAPP}
for details) the `worlds' are to be understood as structures or patterns in the
underlying quantum state: decoherence, suppressing as it does the
interference between quasiclassically definite states in a
superposition, guarantees that multiple such patterns evolve almost
independently. Applying this to a state like (\ref{cat}) tells us that
we have a world with a live cat and another with a dead cat; applying it
to a state like (\ref{collapsecat}) tells us exactly the same.

Dynamical-collapse theories have the same ontology as the Everett
interpretation; they differ only in dynamics. As such, it seems that
(\ref{collapsecat}) must be interpreted as a many-worlds state, with the
dead-cat `tail' being just as real as the much higher-weight live cat
component.

\emph{How} different are the dynamics? It is often assumed\footnote{See, \egc, 
\cite{albertloewer1990,corderogrw}.} that they are very similar
indeed: the only effect of the collapse mechanism is to damp the
amplitudes of all branches but one, but the branches themselves continue
to evolve normally. Call this the assumption of \emph{quasi-Everettian
dynamics}, or QED. Under unitary dynamics (\ref{cat}) evolves into
something like (schematically)
\[ \nrm\ket{\mbox{Newspapers report `cat lives!'}}\]
\be +\nrm\ket{\mbox{Newspapers report `cat dies!'}};\ee
if QED were true, then, (\ref{collapsecat}) would evolve into
\[ \alpha' \ket{\mbox{Newspapers report `cat lives!'}}\]
\be +\beta'\ket{\mbox{Newspapers report `cat dies!'}}\ee
where again $1>|\alpha'|^2\gg|\beta'|^2>0$.

If QED were true, it would in my view pose a very serious problem for
dynamical collapse theories: a problem very similar, in fact, to the
`empty-wave' or `Everett-in-denial' problem for the de Broglie-Bohm
theory (see \cite{brownwallace} for a presentation of this problem). The
low weight of the `tail' would be empirically undetectable by anyone in
it, and the collapse mechanism would be epiphenomenal.

QED is not true, however: the collapse mechanism has dramatic dynamical
consequences for the tail, as we shall see.

\section{Life in the tails}

It is tempting to think of the GRW collapse mechanism as follows: if a state is initially a
superposition of states localised around points $x$ and $y$, then the
post-collapse state is again such a superposition, just with one of the
localisation peaks greatly magnified in comparison with the other.

This isn't true. If the collapse happens around $x$, say, then it
actually has two effects: the amplitude of the peak around $x$ is
greatly increased relative to the peak around $y$, and the centre of the
peak around $y$ is displaced significantly towards $x$.

It is easy to show this directly: suppose that the initial wave-function
is one dimensional and proportional to
\be \psi(x)=\e{-x^2/2w^2}+ \e{-(x-x_0)^2/2w^2};\ee
that is, suppose it is an equally weighted sum of two Gaussians. The
effect of collapse (assuming that the collapse peak is at x=0) is to
multiply $\psi$ by $\exp(-x^2/2a^2)$; a little algebra gives the result 
\be \psi(x)\longrightarrow \psi'(x)=\e{-x^2/2w'^2}+\e{-(x-
x_0')^2/2w'^2}\e{-x_0^2/a'^2}\ee
where $a'^2=a^2+w^2$, $(1/w'^2)=(1/a^2)+(1/w^2)$, and
$x_0'=x_0 \times a^2/(a^2+w^2)$.
The GRW parameter $a$ is generally taken to be $\sim 10^{-7} m$; on the
assumption that the peaks are much narrower than this (\iec $w\ll a$)
this simplifies approximately to
\be \psi'(x)\simeq \e{-x^2/2w^2}+\e{-(x-
x_0[1-w^2/a^2])^2/2w^2}\e{-x_0^2/a^2}.\ee
So, as well as being shrunk by a factor $\e{-x_0^2/a^2}$, the `tail'
peak has also been displaced a fraction $(w^2/a^2)$ of the distance
towards the collapse centre.\footnote{This should not actually be surprising. The effect of the collapse on
the tail peak is to multiply together two Gaussians with centres a distance $x_0$ apart: one would expect
the resultant function to have a peak somewhere between the two original
peaks.}

The quantitative form of the displacement is dependent on the particular
(Gaussian) form of the peaks used in $\psi(x)$. The overall conclusion,
however, is robust: the effect of collapse on the tail peak is to displace it
towards the collapse centre. From the point of view of an observer in
the tail of a Schr\"{o}dinger-cat state like (\ref{collapsecat}), the
effect is that a particle is `kicked' by the collapse. Furthermore,
since the tail's amplitude is very small, 
any subsequent collapses will almost certainly not prefer (that is, be centred on) the tail, so
all other particles in the tail whose counterparts in the main part of
the wavefunction are spatially separated from them will also be kicked
if they are subject to collapse. 

The `kick' has some interesting consequences for the stability of
matter in the tails. As a simple model, suppose that we work in a mean-field approximation where each particle can be thought of as moving in 
a collective potential set by the other particles; the wavefunction of some multiparticle bound state like an atom or nucleus can then be approximated,
if we neglect the complications caused by the exclusion principle and ignore overal normalization, as a product of individual states. Using as an ansatz the asymptotic behaviour of hydrogen atoms, whose wavefunctions decrease at large distances like 
\be
\psi(x) \propto \e{-|x|/w}
\ee\
then the wavefunction of a collapsed particle in the tail evolves like
\be
\psi(x) \rightarrow \psi'(x) = \e{-r|x|/w} \times \e{-|x-x_0|^2/2a^2}.
\ee
If we assume without loss of generality that the collapse centre is on the positive $x$ axis, restrict the wavefunction to this axis, and differentiate, we get
\be
\frac{\dr \psi'}{\dr x}=\left( \frac{-1}{w} - \frac{x-x_0}{a^2}\right) \psi'(x)
\ee
which takes a maximum at
\be
x_{max}=x_0 - \frac{a^2}{w}
\ee
assuming that $x_0>a>w$. 
That is: the `kick' displaces a bound particle partly outside its parent
compound. If the kick distance is comparable to the actual width $w$ of
the compound, this will at least significantly excite it, and perhaps even
disrupt it. From the above, a rough-and-ready criterion for excitation is 
\be
x_0 > \left( 1+ \frac{a^2}{w^2}\right) \simeq \frac{a^2}{w}
\ee
where in the last step we are assuming $a \gg w$.  Since the normal value of $a$ is taken
to be $\sim 10^{-7}$ m, the kick will cause atomic excitation when $x_0
\sim 10^{-4}$ m and nuclear excitation when $x_0 \sim 1$ m.

To see the practical effects of this, suppose that we prepare a
``Schr\"{o}dinger cat'' state: a macroscopic object of say $10^{27}$ atoms in a superposition
of two locations a couple of metres apart. (This isn't terribly
difficult: Schr\"{o}dinger's original method will do fine.) Within
$10^{-14}$ seconds (assuming a GRW collapse frequency of $10^{-16}
\mathrm{s}$), the first collapse will occur and the amplitude of (say) the second term in the superposition will be drastically reduced relative to the first term.

The macroscopic object contains $\sim 10^{28}$ nucleons and $\sim
10^{28}$ electrons, and $\sim 10^{12}$ of these will undergo dynamical
collapse per second. This will have a completely negligible effect on
the first term in the superposition. The nucleons and electrons represented by
the second (`tail') term, however, will be kicked towards the locations of their counterparts in the first term.

Set aside the kick to the electrons for now (we will see later that it is comparatively unimportant; in any case, in modern `mass density' versions of dynamical-collapse theory, electron collapse is negligibly rare compared to nucleon collapse).
The kick to the
nucleons will kick each nucleon clean outside the
nucleus. Due to the short range of the nuclear force, this will cause
that nucleon to be ejected from the nucleus entirely.

The structure of the nucleus is complicated and not that well understood
quantitatively (and in any case we lack a fully satisfactory version of dynamical-collapse theory for nuclear physics), but in qualitative terms this will lead to (at least)
\begin{enumerate}
\item Gamma radiation as the remnants of the nucleus settle down from
their current highly excited state into the ground state appropriate to
the new number of protons and neutrons in the nucleus.
\item Possible beta or alpha radiation or electron capture, since the remnant nucleus is
probably unstable.
\item If the ejected particle is a neutron, then beta radiation as
it decays.
\end{enumerate}
For instance, a collapse hit on a neutron in the nucleus of a carbon atom
kicks the tail component of that nucleus into a highly excited state of
carbon-11, which (assuming it is not so excited as to break up
altogether) rapidly emits gamma radiation as it relaxes into the ground state and then 
decays via electron capture to boron-11, emitting $\sim 2 \mathrm{MeV}$; the
neutron decays in $\sim$ 10 minutes into a proton-electron-neutrino pair and
emits $\sim 1 \mathrm{MeV}$ in doing so. 

To summarise: if objects in the tails of the wave-function are displaced
by about a metre from the location of their counterpart in the main part
of the wave-function, they become radioactive, with a mean lifetime
equal to the GRW collapse rate --- that is, $\sim 10^{16} \mathrm{s}$,
or about 100 million years. 

\section{Death in the tails}

The decay energies quoted in the previous section --- $\sim 3 \mathrm{MeV}$ --- are
fairly typical  of the energies produced by dynamical-collapse-triggered
decay; as such, a kilogram of matter in the tail which is displaced by
more than a metre or so from its high-weight counterpart will emit
energy at a rate of ($10^{-16}$ decays per second per nucleon $\times$ $\sim
10^{27}$ nucleons per kg $\times$ $\sim 1 \mathrm{MeV}$ per decay)=$\sim
10^{11}$ $\mathrm{MeV}$ per second, or about $10^{-8}$ watts. (This should be
taken as a lower limit as it makes no allowance for gamma radiation when
excited states de-excite.) A large fraction of this energy will be in
the form of highly ionizing beta radiation caused by neutron decay.

This level of radiation will be harmful to living creatures in
the tails. Precise calculations seem inappropriate given the very rough
nature of the estimates used so far. Note, however, that if a living
being (say, Schr\"{o}dinger's unfortunate cat) were to be displaced by
more than a metre or so from its high-weight counterpart, and were to
absorb all of the ionizing radiation emitted by its own radioactive
components (a reasonable order-of-magnitude approximation given the relatively high cross-
section of beta radiation with matter), it would receive a radiation
dose of $\sim 100$ rem per year. This compares very unfavourably to the
Environmental Protection Agency's recommended safe (human) dosage of 100
millirem per year; the EPA quotes 400 rem as the threshold fatal dose. And this is only the radiation exposure from the being's own body, and does not take account of likely radiation from surrounding matter.

(Returning to the effect of the collapse `kick' on the electrons, we can now see that while each such kick ionises a single atom, each nuclear kick will indirectly ionise many atoms through the emission of ionising radiation; hence the health consequences of nucleon kicks dominate those of electron kicks.\footnote{Vaidman~\cite{vaidmandeterminism} analyzes the ionisation effects of GRW electron collapse on the tails; he states that these collapses themselves will very quickly kill observers in the tails, but provides no detailed model. So far as I can see this somewhat exaggerates the health consequences of these ionizations, which amount collectively to the equivalent of a relatively low radiation dose.})

What consequences does this observation have for the measurement
problem? Recall that the dynamical collapse program hoped to establish

\begin{quote}
\textbf{With overwhelmingly high probability, agents will observe
quantum statistics very close to the averages predicted by quantum
mechanics.}
\end{quote}

However, this is threatened by the problem of structured tails. 
If QED had held, we would instead have had

\begin{quote}
\textbf{With overwhelmingly high probability, the overwhelmingly high-weight branch will 
be one in which agents will observe
quantum statistics very close to the averages predicted by quantum
mechanics; however, the weight of a branch is not detectable by any agents, including those who
are in very low-weight, anomalous branches.}
\end{quote}

The failure of QED analysed in this section and the last leads to an
intermediate result:
\begin{quote}
\textbf{With overwhelmingly high probability, agents will either observe
quantum statistics very close to the averages predicted by quantum
mechanics, or in due course die of radiation sickness.}
\end{quote}

Strictly speaking, I suppose that this intermediate result rescues
dynamical-collapse theories from the problem of structured tails and
ensures that they do, after all, solve the measurement problem. It
explains why the scientific community has so far observed statistical results in accord with
quantum mechanics (via the anthropic fact that worlds in which violations were observed are now radioactive deserts).
And it explains why it is rational to act as if the predictions of
quantum mechanics were true (because in those worlds where they turn out
false, we're all doomed anyway). 

However, from a purely sociological viewpoint I suspect that this will
not be deemed adequate by the foundations community. If so, then the
problem of structured tails is real even if its usual description (via
QED) is false. The only recourse that I can see for dynamical-collapse
theorists is then to modify the form of their collapse function. If the
collapse function is taken to have \emph{compact support} (differing
from an exponential at distances of, say, $\sim 10 a$ from the collapse
centre) then the dynamical effect on the main part of the wavefunction
will be completely negligible but the tails will be erased entirely. 

Letting the collapse function have compact support is useless from the
point of view of the problem of bare tails, since the wave-function will
instantaneously evolve to have non-compact support. However, it is the
\emph{structure} of the tails, and not their mere existence, that causes
the problem. Compact support would solve the problem very
straightforwardly\footnote{Whether it would make it easier or harder to generalise dynamical-collapse theories
to relativistic quantum mechanics, I do not know; harder, at a guess, since compact-support functions are less
straightforward to handle than exponentials.} and would spare our low-weight counterparts in the tails from their otherwise grim fate.


\end{document}